\documentclass[12pt]{article}
\pdfoutput=1
\usepackage{graphicx}

\begin{document}

\begin{center}
\Large {\bf  Notes on Derivation of 'Generalized Gravitational Entropy'}
\end{center}

\bigskip
\bigskip

\begin{center}
D.V. Fursaev
\end{center}

\bigskip
\bigskip

\begin{center}
{\it Dubna International University \\
     Universitetskaya str. 19\\
     141 980, Dubna, Moscow Region, Russia\\}
 \medskip
\end{center}

\bigskip
\bigskip

\begin{abstract}
An alternative derivation of generalized gravitational entropy
associated to co-dimension 2 'entangling' hypersurfaces is given. The approach 
is similar to the Jacobson-Myers 'Hamiltonian' method and it does not require
computations on manifolds with conical singularities. It is demonstrated that 
the entangling surfaces should be extrema of
the entropy functional. 
When our approach is applied to Lovelock theories of gravity the generalized entropy 
formula coincides with results derived by other methods.
\end{abstract}

\newpage

\section{Introduction}\label{intr}

There are a mounting number of arguments that the Bekenstein-Hawking entropy 
can be applied not only in case of black hole horizons
but to arbitrary co-dimension 2 surfaces in flat and curved spacetimes. First arguments that this can be done in a consistent way have been presented in the work of the present author \cite{Fursaev:2006ng},\cite{Fursaev:2007sg}.  
If $\cal B$ is a minimal hypersurface in a constant time slice $\Sigma$ of a stationary spacetime $\cal M$ which is a solution to the Einstein theory one can associate to this surface an entropy \cite{Fursaev:2007sg}
\begin{equation}\label{i-1}
S({\cal B})={A({\cal B}) \over 4G}~~,
\end{equation}
where $A({\cal B})$ is the area of $\cal B$.
Equation (\ref{i-1}) has been inspired by the holographic formula \cite{Ryu:2006bv} for computing entanglement entropy in conformal theories with gravity duals. $S({\cal B})$ 
can be interpreted as an entanglement entropy in quantum gravity \cite{Fursaev:2007sg}. A similar
concept of spacetime entanglement 
was discussed in a number of publications, see e.g. \cite{Bianchi:2012ev}, \cite{Myers:2013lva}.

Recently formula (\ref{i-1}) has been also proposed by 
Lewkowycz and Maldacena \cite{Lewkowycz:2013nqa} as a 'generalized gravitational entropy'. The authors of \cite{Lewkowycz:2013nqa} considered a general setup when $\cal M$ is an arbitrary (not necessarily stationary) solution to the Einstein gravity. It was assumed that boundary 
$\partial {\cal M}$ of $\cal M$ has non-contractable circles $S^1$ which are contractable inside $\cal M$ on $\cal B$.  When $\cal B$ is minimal 
in $\cal M$ equation (\ref{i-1}) yields an entropy associated to a density matrix
specified by the given boundary conditions. It was also argued that the above construction
 has an entanglement interpretation.

The Maldacena-Lewkowycz proposal and its extensions to higher derivative gravities attracted a considerable interest \cite{Chen:2013qma}-\cite{Bhattacharyya:2014yga}. The main difficulty here was related to a careful treatment of conical singularities in gravity 
actions \cite{Fursaev:2013fta}. The singularities appeared in \cite{Lewkowycz:2013nqa}
at some steps of computations.

The aim of the present work is to derive 
the generalized gravitational entropy without any use of conical singularities.
Our approach is similar to the Jacobson-Myers 'Hamiltonian' method \cite{Jacobson:1993xs} in a sense
that the entropy appears from a boundary term in the action when one isolates a small
domain around the 'entangling' surface $\cal B$. We 
prove the extremality of
the entropy functional on the
entangling surface and test our approach in Lovelock theories of gravity.

After necessary definitions in Sec. \ref{def} the suggested method is introduced in Sec. \ref{app}. Applications to higher derivative gravities are considered in Sec. \ref{LG}
followed by a brief discussion in Sec. \ref{disc}.

\section{Definitions}\label{def}
\setcounter{equation}0

Entanglement entropy in a quantum gravity, as suggested in
\cite{Fursaev:2007sg}, is specified by the boundary conditions, which imply a holographic nature of the theory. One starts with a class of manifolds $\cal M$
with the boundary condition $\partial {\cal M}={\cal T}$, where $\cal T$ is a $d-1$ dimensional manifold. The entanglement entropy of \cite{Fursaev:2006ng},\cite{Fursaev:2007sg} and the generalized
gravitational entropy of \cite{Lewkowycz:2013nqa} can be defined in terms of 
an 'entanglement' partition function
${\cal Z}[{\cal T}_n]$, where $n=1,2,...$, and ${\cal T}_n$ are boundary manifolds  constructed from  $n$ copies of $\cal T$. 
Construction of
${\cal T}_n$ is similar to a construction of 'replicated' manifolds in a QFT to represent 
quantities like $\mbox{Tr}~\rho^n$, where $\rho$ is a reduced density matrix obtained
by tracing over unobservable states.

The entanglement partition function ${\cal Z}[{\cal T}_n]$ is defined by quantum gravity theory, where bulk geometries ${\cal M}_n$ have the boundary
$\partial {\cal M}_n={\cal T}_n$. One can represent ${\cal Z}[{\cal T}_n]$ in terms
of some integral over 'histories' with above boundary conditions and integration
measure defined by some low-energy action $I[{\cal M}_n]$. 
In a semiclassical approximation 
$\ln {\cal Z}[{\cal T}_n]\simeq - I[\bar{\cal M}_n]$,
where $\bar{\cal M}_n$ realizes a minimum of the action for given boundary conditions, 
and the entropy can be defined as \cite{Fursaev:2007sg}
\begin{equation}\label{of-1}
S=\lim_{n\to 1}(n\partial_n-1)I[\bar{\cal M}_n]~~.
\end{equation}
One first finds the action for integer $n$, assumes that $n$ can be replaced with a continuous parameter, and then goes to  $n=1$. 
This is a common trick used in statistical physics and known as a replica method.

The Maldacena-Lewkowycz approach is to look
for $\bar{\cal M}_n$ as {\it regular} (except some 'harmless' singularities) 
solutions to the corresponding low-energy gravity
equations with the condition $\partial \bar{\cal M}_n={\cal T}_n$.
It is assumed that $\bar{\cal M}=\bar{\cal M}_{1}$ is one of solutions for 
standard boundary conditions $\partial \bar{\cal M}={\cal T}$. In the alternative approach
\cite{Fursaev:2007sg} $\bar{\cal M}_n$ are allowed to have conical singularities. The idea 
of \cite{Fursaev:2007sg} is that gravity actions may have minima  
on such backgrounds if ${\cal T}_n$ have conical singularities.

For the purposes of the present paper we follow \cite{Lewkowycz:2013nqa}.  
Here 
the boundary manifolds ${\cal T}$ are required to have non-contractable circles $S^1$. One can introduce a coordinate $\tau$ along the circles with the period $2\pi$. The boundary manifold ${\cal T}_n$ for the partition function in the replica method 
is glued {\it smoothly} from $n$ copies of ${\cal T}_n$ 
such that $\tau$ has the period $2\pi n$. It is required that 
${\cal T}$ and ${\cal T}_n$ are boundaries of manifolds where
$S^1$ can be contracted in the bulk. A simple example is 
the case of a black hole instanton, where $\cal M$ is a solid hypertorus for 
$\partial {\cal M}=S^1\times S^{d-1}$.

In the rest of the paper we use the following notations: $R_{\mu\nu\lambda\rho}$ is the Riemann tensor of a $d$ dimensional manifold $\cal M$, $\cal M$ has the Euclidean signature. The Greek indexes run from 1 to $d$. $\partial {\cal M}$ is (either external 
or internal) boundary of $\cal M$, $K^a_b$ is the extrinsic curvature tensor of $\partial {\cal M}$. The Latin indexes $a,b,c,d$ run from 1 to $d-1$. A tensor $R_{abcd}$ 
on $\partial {\cal M}$ is
a projection of the Riemann tensor of $\cal M$ on a space tangent to $\partial {\cal M}$.
$\cal B$ is an 'entangling' co-dimension 2 hypersurface in $\cal M$. We use a unit complex
vector constructed from two normal vectors to $\cal B$ and define 
the corresponding complex extrinsic curvature $k^i_j$. The Riemann tensor defined by the metric
of $\cal B$ is denoted as $\hat{R}_{ijkl}$.  The Latin indexes $i,j,k,l$ run from 1 to $d-2$.
The operation $[\mu_1,...,\mu_p]$ denotes totally antisymmetric combination of $p$ 
indexes (accompanied by the factor $1/p!$).

\section{A novel derivation of the generalized entropy}\label{app}
\setcounter{equation}0

To present the method we start with the Einstein gravity. 
Let $\bar{\cal M}_n$ be a solution to gravity equations for corresponding boundary conditions
${\cal T}_n$. By following \cite{Lewkowycz:2013nqa} we assume that a $Z_n$ symmetry of boundary conditions
${\cal T}_n$ (permutations of replicas) is extended to the bulk.
Let  ${\cal B}_n$ be a surface of fixed points in $\bar{\cal M}_n$ (points which do not move 
under the $Z_n$ symmetry).  Maldacena and Lewkowycz \cite{Lewkowycz:2013nqa} interpret 
${\cal B}_n$ as a world-sheet of a cosmic string (brane) and derive conditions on ${\cal B}_n$ from 
a regularity condition on the geometry around a cosmic string. We consider sets of solutions $\bar{\cal M}_n$ and 
corresponding surfaces ${\cal B}_n$ but do not write the index  $n$ explicitly, for a while.

A 'cosmic string' action on ${\cal B}$ can be inferred immediately from 
the gravity action on $\bar{\cal M}$. The idea is the following.  Consider a small neighbourhood 
${\cal N}_\epsilon$ around ${\cal B}$, where the 
metric, according to \cite{Lewkowycz:2013nqa}, behaves as 
\begin{equation}\label{eg-1}
ds^2\simeq r^2d\tau^2+n^2dr^2+
\left(\gamma_{ij}(v)+2r^nc^{1-n}(\cos\tau~k^{(1)}_{ij}(v)+\sin \tau~k^{(2)}_{ij}(v))\right)
dv^idv^j~~.
\end{equation}
Here $0<\tau \leq 2\pi n$, $0<r \leq \epsilon$, $c$ is a dimensional constant, $\gamma_{ij}(v)$ is a metric on $\cal B$, and 
$k^{(p)}_{ij}(v)$ are two extrinsic curvatures of $\cal B$.  Metric (\ref{eg-1}) is invariant under 
shifts $\tau\to \tau +2\pi$. This property ensures the $Z_n$ symmetry. 
The given metric does not have conical singularities in the
($r,\tau$) part, and the geometry is regular for all natural $n$. 
For arbitrary $n$ (for example, $n$ slightly larger than 1)
(\ref{eg-1}) is not regular at $r=0$ due to terms 
with extrinsic curvatures.
Components of the Ricci tensor have power-law divergences if $1 < n < 2$. 
In this work we adopt the point of view of \cite{Lewkowycz:2013nqa} 
that the singularity due to the extrinsic curvature terms is 'harmless' in a 
sense it disappears in components of the Einstein tensor
if $\cal B$ obeys certain conditions. For the Einstein gravity this condition is that 
$\cal B$ is minimal (extrinsic curvatures $k^{(p)}_{ij}$ have vanishing traces). 

In coordinates (\ref{eg-1}) 
the boundary of the neighbourhood is chosen to be located at $r=\epsilon$.
The gravity action on $\bar{\cal M}$
is decomposed on the action on ${\cal N}_\epsilon$ and the action on 
$\bar{\cal M}/{\cal N}_\epsilon$. It is assumed that a necessary boundary 
term with the extrinsic curvature on the boundary of ${\cal N}_\epsilon$ is included
in the actions to have a well-posed variational problem. 
In the limit $\epsilon \to 0$ the action on ${\cal N}_\epsilon$  
can be interpreted as a 'cosmic string' action $I_{\mbox{\tiny{str}}}$,
\begin{equation}\label{on-4}
I_{\mbox{\tiny{str}}}[{\cal B}]=\lim_{\epsilon \to 0} I[{\cal N}_\epsilon]=-A({\cal B})/(4G)~~,
\end{equation}
\begin{equation}\label{on-5}
I[{\cal N}_\epsilon]=-{1 \over 16\pi G}\int_{{\cal N}_\epsilon}\sqrt{g}d^dx ~R
-{1 \over 8\pi G}\int_{\partial {\cal N}_\epsilon}\sqrt{h}d^{d-1}y ~K~~.
\end{equation}
To get (\ref{on-4}) from
(\ref{on-5}) one should take into account that the extrinsic curvature tensor 
$K^\mu_\nu$
of ${\cal N}_\epsilon$ has a singular component $K_\tau^\tau=1/(n\epsilon)$.
This singularity is compensated by the factor $\epsilon$ in the integration measure.
The bulk part of $I[{\cal N}_\epsilon]$ vanishes in this limit. The 'cosmic string' has the negative tension $-1/(4G)$.
Thus, the use of this terminology is only for an analogy, not for drawing any physical consequences. We also note that our definition of the  'string' differs from 
how it was originally introduced in \cite{Lewkowycz:2013nqa},\cite{Dong:2013qoa}.
Our 'string' has a finite tension when $n\to 1$.

In the limit $\epsilon \to 0$ one can write
\begin{equation}\label{on-2}
I[\bar{\cal M}]=I[\bar{\cal M}^c]+I_{\mbox{\tiny{str}}}[{\cal B}]~~,
\end{equation}
where $I[\bar{\cal M}^c]$ is an action on 
a manifold $\bar{\cal M}^c=\bar{\cal M}/{\cal B}$, where $\cal B$ is removed.
Variation of (\ref{on-2}) over the metric yields the Einstein equations 
outside ${\cal B}$. These are the vacuum equations if the matter is absent. 

Variation of the 'string action' is easy to understand at a small but finite
$\epsilon$ (at a finite string thickness). There are non-trivial 
variations on the boundary ${\cal N}_\epsilon$ due to the boundary terms 
in the gravity action on $\bar{\cal M}/{\cal N}_\epsilon$ and in the 'string'
domain ${\cal N}_\epsilon$. This yields equations
\begin{equation}\label{on-3}
(K^{\mu\nu}-h^{\mu\nu}K)_+=-(K^{\mu\nu}-h^{\mu\nu}K)_-~~.
\end{equation}
Here $(K_+)^\mu_\nu$ and $(K_-)^\mu_\nu=K^\mu_\nu$ are the extrinsic curvatures of 
${\cal N}_\epsilon$ in $\bar{\cal M}/{\cal N}_\epsilon$ and ${\cal N}_\epsilon$, respectively. The left hand side comes out from the 'gravity part' and the right hand side
from the 'string'. The r.h.s. of (\ref{on-3}) can be interpreted as a 'stress-energy tensor'
of the 'string'.
Equations (\ref{on-3}) are identities since the division on the gravity and 'string' parts is artificial. 

From now on the index $n$ is restored. Before applying formula (\ref{of-1}) we discuss variation of $I[\bar{\cal M}_n]$ 
over $n$. We use the same arguments as in \cite{Lewkowycz:2013nqa} and consider 
$I[\bar{\cal M}_n]$ as some integrals at continuous $n$.  

Let us start with decomposition  (\ref{on-2}). For $I[\bar{\cal M}^c_n]$ 
extrapolation to continuous $n$
does not pose a problem since
a small domain near ${\cal B}_n$ is excluded. Variation over $n$ can be written as
\begin{equation}\label{on-6}
\partial_nI[\bar{\cal M}^c_n]=\partial_n ^{\mbox{\tiny{int}}}I[\bar{\cal M}^c_n]
+\partial_n ^{\mbox{\tiny{bulk}}}I[\bar{\cal M}^c_n]+
\partial_n ^{\mbox{\tiny{boun}}}I[\bar{\cal M}^c_n]~~.
\end{equation}
The operation $\partial_n ^{\mbox{\tiny{int}}}$ means a change of the upper limit in the integrals in $\tau$, when the integrand itself is fixed. This is equivalent to 
changing the number of replicas or the periodicity of $\tau$. Variations 
$\partial_n ^{\mbox{\tiny{bulk}}}$, $\partial_n ^{\mbox{\tiny{boun}}}$ take into account, respectively, change of metrics in the bulk and on the boundaries of ${\cal M}^c_n$ 
(when the period of $\tau$ is fixed.)
Variation of the string action can be written as 
\begin{equation}\label{on-9}
\partial_n I_{\mbox{\tiny{str}}}[{\cal B}_n]=
\partial_n ^{\mbox{\tiny{metr}}}I_{\mbox{\tiny{str}}}[{\cal B}_n]+
\partial_n ^{\mbox{\tiny{pos}}}I_{\mbox{\tiny{str}}}[{\cal B}_n]~~,
\end{equation}
where $\partial_n ^{\mbox{\tiny{metr}}}$ corresponds to the variation of the
metric of ${\cal B}_n$, while $\partial_n ^{\mbox{\tiny{pos}}}$ takes into account
change in the position of ${\cal B}_n$ under fixed metric. If ${\cal B}_n$  is a minimal
surface the change of the position does not change the string action in the leading order.

We need variations at $n=1$. Since the $Z_n$ symmetry is implied
\begin{equation}\label{on-7}
\lim_{n\to 1}\partial_n ^{\mbox{\tiny{int}}}I[\bar{\cal M}^c_n]=I[\bar{\cal M}^c]~~,
\end{equation}
where $\bar{\cal M}^c=\bar{\cal M}^c_1$. Equation (\ref{on-7}) is easy to understand when
the metric does not depend on $\tau$. 
One also has
\begin{equation}\label{on-8}
\lim_{n\to 1}\partial_n ^{\mbox{\tiny{bulk}}}I[\bar{\cal M}^c_n]=0~~.
\end{equation}
The action has an extremum on $\bar{\cal M}^c_n$.

Since the metric on the external boundary is fixed one should care about 
variations on the internal boundary of $\bar{\cal M}^c_n$. 
The latter are compensated by the variations of the string action,
\begin{equation}\label{on-10}
\partial_n ^{\mbox{\tiny{metr}}}I_{\mbox{\tiny{str}}}[{\cal B}_n]+\partial_n ^{\mbox{\tiny{boun}}}I[\bar{\cal M}^c_n]=0~~.
\end{equation}
Eq. (\ref{on-10}) is ensured by gravity equations (\ref{on-3}. There is a subtle point here. Variations of the parameter $n$ in the 
metric under fixed periodicity of $\tau$ result in conical singularities in 
(\ref{eg-1}), see \cite{Lewkowycz:2013nqa}. Thus, (\ref{on-10}) is satisfied up to terms
$O(n-1)$. There is no real cosmic string to support the singularity. Therefore, (\ref{on-10}) holds only in the limit $n\to 1$, 
which is enough for our purposes.

By taking into account equations (\ref{on-9}), (\ref{on-7}), (\ref{on-8}), (\ref{on-10}) one finds
\begin{equation}\label{on-11}
\lim_{n\to 1}\partial_nI[\bar{\cal M}_n]=I[\bar{\cal M}]+\lim_{n\to 1}\partial_n ^{\mbox{\tiny{pos}}}I_{\mbox{\tiny{str}}}[{\cal B}_n]~~,
\end{equation}
\begin{equation}\label{on-12}
S=-I_{\mbox{\tiny{str}}}[{\cal B}]+\lim_{n\to 1}\partial_n ^{\mbox{\tiny{pos}}}I_{\mbox{\tiny{str}}}[{\cal B}_n]~~.
\end{equation}
The Bekenstein-Hawking formula (\ref{i-1}) follows from (\ref{on-12}) if one uses
(\ref{on-4}) and assumes that $\cal B$ is a minimal surface ($\partial_n ^{\mbox{\tiny{pos}}}I_{\mbox{\tiny{str}}}[{\cal B}_n]=0$).

\section{Entropy formula in the Lovelock gravity}\label{LG}
\setcounter{equation}0

From Eqs. (\ref{on-4}), (\ref{on-12}) the generalized entropy can be written 
as 
\begin{equation}\label{on-12-a}
S=-I_{\mbox{\tiny{str}}}[{\cal B}]=-\lim_{\epsilon \to 0} I[{\cal N}_\epsilon]~~,
\end{equation}
and it is a pure boundary term.
This equality does not require that the theory is of the Einstein 
form. It can be also applied to higher derivative gravities provided that: 
{\it a}) the action functional admits boundary terms which insure well-posed 
variational procedure (normal derivatives of the metric
variations do not appear on the boundary); {\it b}) the theory admits solutions 
$\bar{\cal M}_n$ for the given boundary conditions
$\partial \bar{\cal M}_n={\cal T}_n$ with the required $Z_n$ symmetry; 
{\it c}) 
${\cal B}$ is an extremum of $I_{\mbox{\tiny{str}}}[{\cal B}]$ (remember that this condition 
eliminates the last term in the r.h.s. of (\ref{on-12})); 
{\it d}) singularities
of the solutions near fixed point surfaces ${\cal B}_n$ are 'harmless' in a sense that 
divergences in the gravity equations (in the higher curvature analogue of the Einstein
tensor) are eliminated by certain conditions on ${\cal B}_n$, and these conditions at $n=1$ 
are equivalent to equations which follow from requirement ({\it c}).

An example of a higher derivative gravity, where ({\it a}) is satisfied, is the Lovelock theory
\begin{equation}\label{lg-1}
I_L[{\cal M}]= -\sum_m c_m\left(\int_{\cal M}{\sqrt g}d^dx ~L_m+
\int_{\partial {\cal M}}{\sqrt h}d^{d-1}y ~B_m\right)~~.
\end{equation}
Here $c_m$ are some coefficients, $c_1>0$, and
\begin{equation}\label{lg-2}
L_m={(2m)! \over 2^m} R^{[\mu_1\nu_1}_{[\mu_1\nu_1}R^{\mu_2\nu_2}_{\mu_2\nu_2}....
R^{\mu_m\nu_m]}_{\mu_m\nu_m]}~~,
\end{equation}
\begin{equation}\label{lg-3}
B_m={(2m)! \over 2^{m-1}} \sum_{p=0}^{m-1} d_{m,p}~ K^{[a_1}_{[a_1}K^{a_2}_{a_2}....K^{a_{2p+1}}_{a_{2p+1}}
R^{b_1c_1}_{b_1c_1}R^{b_2c_2}_{b_2c_2}....
R^{b_{m-p-1}c_{m-p-1}]}_{b_{m-p-1}c_{m-p-1}]}~~,
\end{equation}
\begin{equation}\label{lg-3a}
d_{m,p}={ (m-1)! 2^{3p} p!\over (m-p-1)!(2p+1)! }~~.
\end{equation}
It is implied that $R^{\mu\nu}_{\mu'\nu'}=R^{\mu\nu}_{~~~\mu'\nu'}$, $R^{ab}_{a'b'}=R^{ab}_{~~~a'b'}$.
Curvatures in the r.h.s. of (\ref{lg-3}) are taken on $\cal B$.
We use the form of the boundary term (\ref{lg-3}) given in \cite{Neiman:2013ap}. 

Consider the Lovelock action
in a small domain ${\cal N}_\epsilon$,  where the 
metric behaves as in (\ref{eg-1}).
As earlier, we place the boundary 
$\partial {\cal N}_\epsilon$ at $r=\epsilon$. The 'string action' in this theory is
determined  by the boundary terms on $\partial {\cal N}_\epsilon$.

We need to study boundary terms in (\ref{lg-1}) in the limit $\epsilon\to 0$.
Since the only singular component of $K^a_b$ is $K^\tau_\tau=1/\epsilon$ one can easily see that
$B_m\sim 1/\epsilon$ at $\epsilon\to 0$. The singular terms can be easily extracted
from (\ref{lg-3}):
\begin{equation}\label{lg-4}
K^{[a_1}_{[a_1}....K^{a_{2p+1}}_{a_{2p+1}}
R^{b_1c_1}_{b_1c_1}....
R^{b_{m-p-1}c_{m-p-1}]}_{b_{m-p-1}c_{m-p-1}]} \simeq {2p+1 \over 2m-1}K^\tau_\tau 
K^{[i_1}_{[i_1}...K^{i_{2p}}_{i_{2p}}
R^{j_1k_1}_{j_1k_1}...
R^{j_{m-p-1}k_{m-p-1}]}_{j_{m-p-1}k_{m-p-1}]}~~.
\end{equation}
The factor $2p+1$ in the r.h.s. of (\ref{lg-4}) appears since a pair of upper and lower $\tau$ indexes take $2p+1$ positions, $2m-1$ in the denominator results from the normalization factor in the operator $[...]$. The indexes $i,j,k$ enumerate components of the curvature tensors in the directions tangent 
to $\cal B$. 

It is convenient to introduce complex extrinsic curvatures of $\cal B$
\begin{equation}\label{lg-5}
k_{ij}=\frac 12(k^{(1)}_{ij}-ik^{(2)}_{ij})~~,~~\bar{k}_{ij}=k_{ij}^*~~
\end{equation}
and use the relation, which follows from (\ref{eg-1}) at $n=1$, 
\begin{equation}\label{lg-5a}
K_{ij}=e^{i\tau}k_{ij}+e^{-i\tau}\bar{k}_{ij}~~
\end{equation}
(we assume $n=1$ for the r.h.s. of (\ref{on-12-a})).
Integration over the $\tau$ coordinate can be easily done, 
$$
\int_0^{2\pi}d\tau K^{[a_1}_{[a_1}....K^{a_{2p+1}}_{a_{2p+1}}~
R^{b_1c_1}_{b_1c_1}....
R^{b_{m-p-1}c_{m-p-1}]}_{b_{m-p-1}c_{m-p-1}]} \simeq 
$$
\begin{equation}\label{lg-6}
{2\pi \over \epsilon}{(2p+1) \over (2m-1)}{(2p)! \over p!} 
k^{[i_1}_{[i_1}....k^{i_{p}}_{i_{p}}
\bar{k}^{i_{p+1}}_{i_{p+1}}....\bar{k}^{i_{2p}}_{i_{2p}}
~R^{j_1k_1}_{j_1k_1}....
R^{j_{m-p-1}k_{m-p-1}]}_{j_{m-p-1}k_{m-p-1}]}~~.
\end{equation}
The factor $(2p)!/p!$ in the r.h.s. of (\ref{lg-6}) counts the number of ways 
when $p$ $k$-curvatures (or $\bar k$-curvatures) appear from $2p$ $K$-curvatures.

When (\ref{lg-6}) is used in the boundary term (see (\ref{lg-3})) one comes to the action  
\begin{equation}\label{lg-10}
\lim_{\epsilon \to 0} I_L[{\cal N}_\epsilon]=-4\pi \sum_m  m c_m \hat{I}_m[{\cal B}]~~,
\end{equation}
\begin{equation}\label{lg-11}
\hat{I}_m[{\cal B}]=\int_{\cal B}\hat{L}_{m-1}~~,
\end{equation}
$$
\hat{L}_{m-1}={(2(m-1))! \over 2^{m-1}}\sum_{p=0}^{m-1}{2^{3p}(m-1)! \over p!(m-p-1)!} 
$$
\begin{equation}\label{lg-8}
~k^{[i_1}_{[i_1}....k^{i_{p}}_{i_{p}}
\bar{k}^{i_{p+1}}_{i_{p+1}}....\bar{k}^{i_{2p}}_{i_{2p}}
~R^{j_1k_1}_{j_1k_1}....
R^{j_{m-p-1}k_{m-p-1}]}_{j_{m-p-1}k_{m-p-1}]}~~.
\end{equation}
One can now see that the last equation (\ref{lg-8}) is of the Lovelock form on $\cal B$,
\begin{equation}\label{lg-12}
\hat{L}_{m-1}=
{(2(m-1))! \over 2^{m-1}} \hat{R}^{[i_1j_1}_{[i_1j_1}\hat{R}^{i_2j_2}_{i_2j_2}....
\hat{R}^{i_{m-1}j_{m-1}]}_{i_{m-1}j_{m-1}]}~~.
\end{equation}
Eqs. (\ref{lg-11}), (\ref{lg-8}) follow from (\ref{lg-12}) if one uses in (\ref{lg-12})
the Gauss-Codazzi equations on $\cal B$
\begin{equation}\label{lg-9}
\hat{R}^{j_1j_2}_{i_1i_2}=R^{j_1j_2}_{i_1i_2}+2(k^{j_1}_{i_1}\bar{k}^{j_2}_{i_2}+
\bar{k}^{j_1}_{i_1}k^{j_2}_{i_2}-k^{j_1}_{i_2}\bar{k}^{j_2}_{i_1}-
\bar{k}^{j_1}_{i_2}k^{j_2}_{i_1})
~~.
\end{equation}
Factor $(m-1)! /(p!(m-p-1)!)$ yields a number of ways to pick up $p$  $k\bar{k}$-pairs.
Multiplier $2^{3p}$ takes into account factor 2 in the r.h.s. of (\ref{lg-9}) and the fact that each Riemann curvature in (\ref{lg-9}) produces 4 $k\bar{k}$-pairs.

We come to the following formula
of the generalized entropy associated to the surface $\cal B$:
\begin{equation}\label{lg-13}
S=4\pi \sum_m  m c_m \hat{I}_m[{\cal B}]~~.
\end{equation}
In a context of the holographic entanglement entropy 
(\ref{lg-13}) has been suggested in \cite{Hung:2011xb}.
In case of the Gauss-Bonnet gravity this entropy formula has been obtained 
by different methods: in \cite{Fursaev:2013fta} by using regularized conical singularity
method and in  \cite{Chen:2013qma},\cite{Bhattacharyya:2013jma} from the requirement 
of regularity of the geometry around the 'cosmic string'. For Lovelock 
gravities  (\ref{lg-13}) was also derived in \cite{Dong:2013qoa}.

There are arguments \cite{Dong:2013qoa}  that the Lovelock gravity satisfies condition ({\it d}) if ${\cal B}$ is an extremum of (\ref{lg-13}). A careful study 
of this property in case of the Gauss-Bonnet gravity can be found in \cite{Bhattacharyya:2014yga}.

\section{Discussion}\label{disc}

We presented a sketch of arguments which may support the Maldacena-Lewkowycz proposal \cite{Lewkowycz:2013nqa} when the low-energy gravity action has higher derivatives.
We have not yet emphasized but implied that this construction should be also applicable
to holographic entanglement entropy. In this case $\cal B$ is a holographic entangling surface and the background manifold $\cal M$ is a solution to an AdS gravity.

Our arguments (and, perhaps, other derivations of the generalized gravitational 
entropy) cannot be considered as a sort of a mathematical proof. One should demonstrate
that gravity solutions for given boundary conditions for each value of the replica parameter $n$ do exist and obey condition ({\it d}) formulated in sec. \ref{LG}. If this is the case the generalized entropy can be derived as a limiting value of a boundary term
in the action. The derivation is self-consistent if the entangling surface is an extremum 
of the entropy functional. 

One should mention that ({\it d}) may not be respected in arbitrary higher derivative gravities \cite{Bhattacharyya:2014yga}. 

In contrast to \cite{Lewkowycz:2013nqa}  the approach of \cite{Fursaev:2007sg} operates
with singular geometries. By the construction, the bulk manifolds ${\cal M}_n$ in \cite{Fursaev:2007sg} are replicas of ${\cal M}_1$ with conical singularities at $\cal B$. 
The two ways, \cite{Fursaev:2007sg} and \cite{Lewkowycz:2013nqa}, lead to 
the same entanglement entropy but yield different results for the Renyi entropies. It may happen that the two approaches compliment each other and the
choice between them is determined by studying for which 
background the gravity action has a least value.

\bigskip
\bigskip
\bigskip

\noindent
{\bf Acknowledgement}

\bigskip

The author acknowledges a support from RFBR grant 13-02-00950.
The author is also grateful to J. Camps, A. Patrushev, and S. Solodukhin for comments which helped to correct the discussion of singularities in (\ref{eg-1}).


\newpage

\begin{thebibliography}{}

\bibitem{Fursaev:2006ng} D.V. Fursaev, {\it Entanglement Entropy in Critical Phenomena and Analogue Models of Quantum Gravity}, Phys. Rev. {\bf D73} (2006) 124025
e-Print: hep-th/0602134.

\bibitem{Fursaev:2007sg} D.V. Fursaev, {\it Entanglement entropy in quantum gravity and the Plateau problem}, Phys. Rev. {\bf D77} (2008) 124002,
e-Print: arXiv:0711.1221 [hep-th].
 
\bibitem{Ryu:2006bv} S. Ryu and T. Takayanagi, 
{\it Holographic Derivation of Entanglement Entropy from AdS/CFT}, 
Phys. Rev. Lett. {\bf 96} (2006) 181602,
e-Print: hep-th/0603001.

\bibitem{Bianchi:2012ev} E. Bianchi, R.C. Myers, {\it On the Architecture of Spacetime Geometry}, 
e-Print: arXiv:1212.5183 [hep-th]. 

\bibitem{Myers:2013lva} R.C. Myers, R. Pourhasan, M. Smolkin, 
{\it On Spacetime Entanglement}, JHEP {\bf 1306} (2013) 013, e-Print: arXiv:1304.2030 [hep-th].

\bibitem{Lewkowycz:2013nqa} A.~Lewkowycz and J.~Maldacena,
{\it Generalized gravitational entropy}, JHEP {\bf 1308} (2013) 090,
 e-Print: arXiv:1304.4926 [hep-th].

\bibitem{Chen:2013qma} Bin Chen and Jia-ju Zhang, {\it Note on generalized gravitational entropy in Lovelock gravity}, JHEP {\bf 1307} (2013) 185, 
e-Print: arXiv:1305.6767 [hep-th].

\bibitem{Bhattacharyya:2013jma} A. Bhattacharyya, A. Kaviraj, A. Sinha
{\it Entanglement entropy in higher derivative holography}, JHEP {\bf 1308} (2013) 012,
e-Print: arXiv:1305.6694 [hep-th].

\bibitem{Bhattacharyya:2013gra} A. Bhattacharyya, M. Sharma, A. Sinha, {\it On generalized gravitational entropy, squashed cones and holography}, JHEP {\bf 1401} (2014) 021, 
e-Print: arXiv:1308.5748 [hep-th]. 

\bibitem{Dong:2013qoa} Xi Dong, {\it Holographic Entanglement Entropy for General Higher Derivative Gravity}, JHEP {\bf 1401} (2014) 044, e-Print: arXiv:1310.5713 [hep-th].

\bibitem{Camps:2013zua} J. Camps, {\it Generalized Entropy and Higher Derivative Gravity},
JHEP {\bf 1403} (2014) 070, e-Print: arXiv:1310.6659 [hep-th].

\bibitem{Bhattacharyya:2014yga} A. Bhattacharyya, M. Sharma, {\it On entanglement entropy functionals in higher derivative gravity theories}, e-Print: arXiv:1405.3511 [hep-th].

\bibitem{Fursaev:2013fta} D.V. Fursaev, A. Patrushev, S.N. Solodukhin,
{\it Distributional Geometry of Squashed Cones}, Phys. Rev. {\bf D88} (2013) 4, 044054, 
e-Print: arXiv:1306.4000 [hep-th]. 

\bibitem{Jacobson:1993xs} T. Jacobson and  R.C. Myers, {\it Black hole entropy and higher curvature interactions}, 
Phys. Rev. Lett. {\bf 70} (1993) 3684-3687, hep-th/9305016.

\bibitem{Hung:2011xb} L.-Y. Hung, R.C. Myers, M. Smolkin, {\it On Holographic Entanglement Entropy and Higher Curvature Gravity}
JHEP {\bf 1104} (2011) 025, 
e-Print: arXiv:1101.5813 [hep-th].

\bibitem{Neiman:2013ap}  Y. Neiman, {\it The imaginary part of the gravity action and black hole entropy}, JHEP {\bf 1304} (2013) 071, e-Print: arXiv:1301.7041 [gr-qc]. 



\end{thebibliography}
\end{document}